\documentclass[12pt]{iopart}
\usepackage{hyperref,epsfig,graphics}

\expandafter\let\csname equation*\endcsname\relax
\expandafter\let\csname endequation*\endcsname\relax
\usepackage{amssymb,amsmath}

\begin{document}

\title{Turbulent fluctuations around Bjorken flow}
\author{Stefan Floerchinger and Urs Achim Wiedemann}
\address{Physics Department, Theory Unit, CERN, CH-1211 Gen\`eve 23, Switzerland}

\begin{abstract}
We study the evolution of local event-by-event deviations from smooth average fluid dynamic fields,
as they can arise in heavy ion collisions from the propagation of fluctuating initial conditions.
Local fluctuations around Bjorken flow are found to be governed by non-linear equations whose
solutions can be characterized qualitatively in terms of Reynolds numbers. Perturbations at different rapidities 
decouple quickly, and satisfy (after suitable coordinate transformations) 
an effectively two-dimensional 
Navier-Stokes equation of non-relativistic form. We discuss the conditions under which 
non-linearities in these equations cannot be neglected and turbulent behavior is expected to set in. 
\end{abstract}


In recent years, hadronic transverse momentum spectra and their azimuthal dependence with respect to the orientation of the reaction plane have provided tight constraints on the fluid dynamic model of 
ultra-relativistic heavy ion collisions. With the first data from the LHC, and with refined analyses of
RHIC data, higher order flow coefficients $v_3$, $v_4$, $v_5$ and $v_6$ are now starting to complement
the measurements of elliptic flow $v_2$. This provides
access to qualitatively novel features of the collision dynamics. In particular, since event-averaged
initial conditions of heavy ion collisions are by construction symmetric with respect to the reaction plane
at mid rapidity, the recent measurements of non-vanishing odd harmonic coefficients $v_3$, $v_5$ provide 
unambiguous  evidence for the relevance of event-by-event fluctuations in the fluid dynamic evolution. 
Remarkably, at least some of the models currently used to specify initial conditions  
of fluid dynamic simulations can provide naturally for initial event-by-event fluctuations of the phenomenologically
required size~\cite{Alver:2010gr}.
First studies of the fluid dynamic propagation of such geometric initial state fluctuations have 
resulted in marked improvements in the comparison of fluid dynamic simulations with 
data~\cite{Holopainen:2010gz}.

Motivated by these developments, we are studying here the evolution of local event-by-event
deviations from smooth average fluid dynamic fields within the expanding geometry characteristic for
ultra-relativistic heavy ion collisions. We are for instance interested in how a 'primordial'
spectrum of such fluctuations evolves, and which modes of the fluctuation spectrum are damped on 
which time scales. Given that fluid dynamic fluctuations measure deviations from equilibrium, we 
hope to gain in this way novel access to the basic problem of how equilibration can proceed efficiently
in heavy collisions. We also wonder to what extent it is justified to propagate
primordial fluctuations in a linearized ansatz that by the nature of its approximation leaves no room for 
the development of turbulent phenomena. Such an approach is well-motivated in the treatment
of primordial fluctuations in cosmology, but as we shall discuss in the following, the scales and flows
in heavy ion collisions can support a qualitatively different conclusion. 

To be specific, we consider local fluctuations in the fluid dynamic velocity 
$\delta u^\mu$ and energy density $\delta \epsilon$ on top of average fields $ \bar u^\mu $
and $\bar \epsilon$ that satisfy Bjorken's scaling solution. We work in light cone coordinates,
where $\bar u^\mu=(1,0,0,0)$ and $\bar \epsilon = \epsilon_{Bj}(\tau_0) \left(\tau_0/\tau \right)^{4/3}$. 
In general, one can split at fixed time $\tau$ an arbitrary velocity field $u^\mu = \bar u^\mu + \delta u^\mu$ into
an irrotational part represented by the divergence
\begin{equation}
\vartheta = \partial_1 u^1+ \partial_2 u^2+\partial_y u^y\, ,
\end{equation}
and a solenoidal part represented by the vorticity field 
\begin{equation}
\omega_1 = \tau\, \partial_2 u^y - \frac{1}{\tau} \partial_y u^2, \quad \omega_2 = \frac{1}{\tau} \partial_y u^1 - \tau\, \partial_1 u^y, \quad \omega_3 = \partial_1 u^2 - \partial_2 u^1\, .
\nonumber
\end{equation}
Because of the normalization $u^\mu u_\mu=-1$, there are only 
three independent components $\delta u^j$ which we choose to span the transverse plane and
rapidity, $j=1,2,y$.  The energy density $\epsilon=\bar \epsilon + \delta\epsilon$
can be characterized in terms of temperature deviations from its Bjorken value
\begin{equation}
\hat d = \ln(T/T_\text{Bj}(\tau))\, .
\end{equation}
Within this set-up, we have derived evolution equations for $\delta u^j$ and $\delta \epsilon$
from the relativistic viscous fluid dynamic equations for $u^\mu$ and $\epsilon$.

\begin{figure}
\centering
\setlength{\unitlength}{0.5\textwidth}
\begin{picture}(1,0.7)
\put(0.04,0.04){\includegraphics[width=0.45\textwidth]{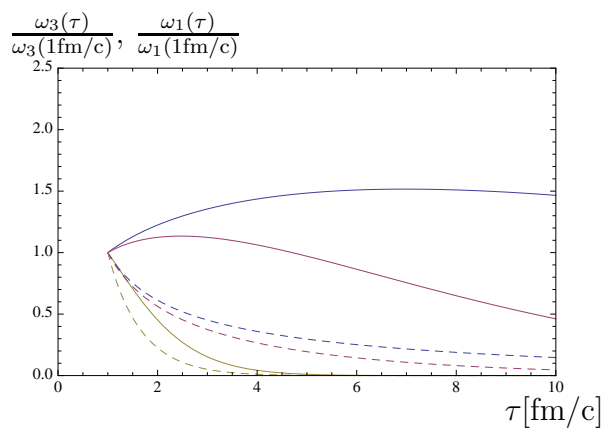}}
\put(0.84,0){$\tau[\text{fm/c}]$}
\put(0,0.64){$\tfrac{ \omega_3(\tau)}{ \omega_3(1\text{fm/c})}$, $\tfrac{ \omega_1(\tau)}{\omega_1(1\text{fm/c})}$}
\end{picture}
\caption{Vorticity amplitudes $ \omega_3(\tau)/ \omega_3(1\text{fm/c})$ (solid lines) and
 $ \omega_1(\tau)/ \omega_1(1\text{fm/c})$ (dashed lines) for wave vectors with $k_2=k_y=0$, and $k_1=5 \,\text{fm}^{-1}$, $k_1=10\, \text{fm}^{-1}$ and $k_1=30\,\text{fm}^{-1}$, respectively. 
 Viscosity has been chosen as $\nu(\tau)/\tau=10^{-3}$ at $\tau=1\,\text{fm/c}$.}
\label{fig:vorticitywave}
\end{figure}

We first comment on the linearized evolution equations for $\delta u^j$, $\delta \epsilon$.
In this case, the equations for $\vartheta$ and $\hat d$ are coupled and describe essentially the propagation of sound (modulo modifications due to the expanding background, see Ref.~\cite{FW11}).
The vorticity modes satisfy a diffusion-type equation of motion that can be solved directly. 
In Fourrier space, denoting by $k_y$ the wave number conjugate to rapidity, 
\begin{equation}
\omega_j(\tau,k_1,k_2,k_y)  =  \omega_j(\tau_0,k_1,k_2,k_y)  \left(\frac{\tau}{\tau_0}\right)^{\tfrac{h_j}{3}} e^{-\tfrac{3\,\nu_0}{4\tau_0^{1/3}} (k_1^2+k_2^2)(\tau^{4/3} -\tau_0^{4/3})+\tfrac{2\,\nu_0 }{3 \tau_0^{1/3}}k_y^2 \left(\tau^{-2/3}-\tau_0^{-2/3}\right)}.
\label{eq:vorticitywave}
\end{equation}
For this solution, we assumed a $\tau$-independent ratio $\eta/s$ that leads to a kinematic viscosity 
 $\nu_0=\eta/(\epsilon+p)=\eta/(sT)$ evaluated at time $\tau_0$. While the transverse vorticity components 
($h_1=h_2 = -2$) fall off like $1/\tau^{2/3}$ for small wave-vectors or small kinematic viscosity $\nu_0$,
the vorticity mode  $\omega_3$ grows algebraically in this region ($h_3 = 1$). Therefore, at least in
a linearized description, there can be initial fluctuations that do not attenuate within the
phenomenologically relevant time scales of heavy ion collisions, see Fig.\ \ref{fig:vorticitywave}.

In general, a linearized formalism applies if perturbations are small  $\delta \epsilon/\bar \epsilon \ll 1$, $d \ll 1$ 
{\it and} if Reynolds numbers are small, 
\begin{equation}
\text{Re} = \frac{u_T\, l\, (\epsilon + p)}{\eta} = \frac{u_T\, l\, s\, T}{\eta}.
\label{eq:Reynoldsnumber}
\end{equation}
Here $u_T$ is a characteristic velocity of fluctuations in the transverse direction and $l$ is the typical length scale over which it changes significantly. (For motion in the rapidity direction, there is a different Reynolds number~\cite{FW11}.)
If the Reynolds number is large, then turbulent flow sets in and a linearized formalism fails. In this case, 
the smallness of perturbations around an expanding background still leads to the important simplification
that the fluid can be treated as compression-less, $\vartheta=0$. This condition does not mean that there are no sound waves present, but that the coupling between sound and turbulence becomes negligible. The formal criterion is a small Mach number
\begin{equation}
\text{Ma} = \frac{\sqrt{u_1 u^1 + u_2 u^2+ u_y u^y}}{c_S} \ll 1\, ,
\end{equation}
where $c_S$ is the velocity of sound. It is
conceivable that the condition $\text{Ma} \ll 1$ for compression-less turbulent flow is realized in heavy ion 
collisions. 

In terms of a rescaled time  $t = 3 \tau^{4/3}/(4 \tau_0^{1/3})$, temperature field
$d=(\tau_0/\tau)^{2/3} \hat d$ and  velocity $v_j = (\tau_0/\tau)	^{1/3} u_j$, we find that the non-linear fluid
equations ($j=1,2,y$) take the form
\begin{equation}
\partial_t v_j + \sum_{m=1}^2 v_m \partial_m v_j  \;+\, \frac{1}{\tau^2} v_y\partial_y v_j +\partial_j d
-\nu_0 \left( \partial_1^2 + \partial_2^2  \;+\, \frac{1}{\tau^2} \partial_y^2 \right) v_j =0
\label{eq:NavierStokes}
\end{equation}
with the solenoidal constraint $
\partial_1 v_1 + \partial_2 v_2 +  \frac{1}{\tau^2}\partial_y v_y=0$.
For late times, this becomes effectively a two-dimensional Navier-Stokes equation of non-relativistic form
for a two-dimensional compression-less fluid. Therefore, the development of turbulent flow in heavy ion 
collisions can be discussed on the basis of an equation about which much is known already. 
 In particular, for the three-dimensional Navier-Stokes equation, one knows that kinetic energy cascades 
 in the case of turbulence from large structures in space to finer and finer ones where it is eventually dissipated. 
In contrast, in two dimensions, turbulent kinetic energy is subject to an inverse cascade from microscopic to 
more and more macroscopic structures. 
This plays a role for turbulent phenomena in the essentially two-dimensional
layer of the Earth's atmosphere and it prompts us to wonder whether (\ref{eq:NavierStokes}) allows
for a scale amplification mechanism of some fluctuating modes in heavy ion collisions. 

\begin{figure}
\centering
\setlength{\unitlength}{0.52\textwidth}
\begin{picture}(1,0.8)
\put(0.04,0.07){\includegraphics[width=0.5\textwidth]{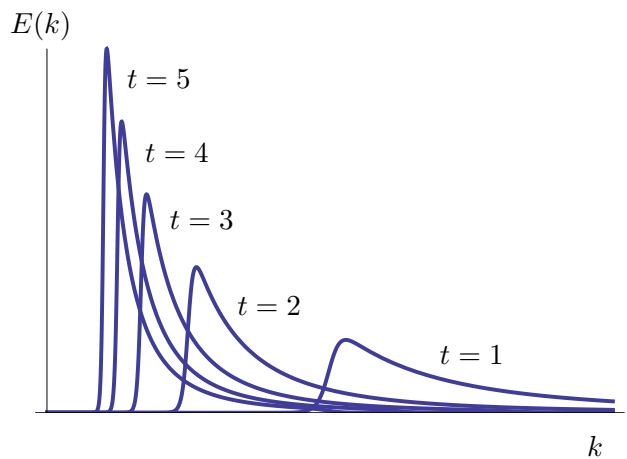}}
\put(0.94,0){\small $k$}
\put(0,0.69){{\small $E(k)$}}
\put(0.7,0.15){\small $t=1$}
\put(0.37,0.23){\small $t=2$}
\put(0.26,0.37){\small $t=3$}
\put(0.22,0.48){\small $t=4$}
\put(0.19,0.60){\small $t=5$}
\end{picture}
\caption{Illustration of turbulent kinetic energy as a function of the wave-number for freely decaying turbulence according to Batchelors scaling theory.}
\label{fig:Batchelor}
\end{figure}

One way to characterize the inverse cascade is to study how the kinetic
energy in a fluid,
\begin{equation}
\lambda^2=\frac{1}{2}\langle v_1^2 + v_2^2 \rangle = \int_0^\infty d k\; E(k),
\end{equation}
is distributed over wave vectors $E(k)$ as a function of time. Based on the theory for turbulence developed
by Kolmogorov in three  and Kraichnan in two dimensions, there is in particular a scaling theory for freely
decaying two-dimensional turbulence that leads to  $E(t,k) = \lambda^{3}\, t\, h(k \,\lambda\, t)$, where 
$h(x)$ is conjectured to be universal \cite{Batchelor}. In this case, kinetic energy accumulates at small 
wave-vectors at late times, see Fig.\ \ref{fig:Batchelor}.

The scaling theories of Kolmogorov, Kraichnan and Batchelor address turbulence at very large Reynolds number. 
Estimates of Reynolds numbers realized in heavy ion collisions are usually of the order $\text{Re} \approx s/\eta
= {\cal O}(10)$. This is  too small for applying results from fully developed turbulence, but it is 
sufficiently large to motivate a search for the onset of turbulent phenomena involving fundamental
quantum fields in heavy ion collisions. 
This is a report on work in progress. We are currently exploring possible signatures for the onset of 
turbulent phenomena in measurements of single inclusive hadron spectra and two-particle correlation
functions.

\end{document}